\documentclass{aa}  
\usepackage{graphicx}
\usepackage{graphicx}
\usepackage{amssymb}
\usepackage{subfigure}
\usepackage{natbib}
\usepackage{tabularx}
\usepackage{longtable}
\usepackage{lscape}
\bibpunct{(}{)}{;}{a}{}{,} 
\begin{document}
   \title{An Outbursting Protostar of the FU~Orionis type in the Cygnus OB7 Molecular 
          Cloud\thanks{Based on observations collected at the Byurakan Astrophysical 
          Observatory (Armenia), Special Astrophysical Observatory (Russia), 
          German-Spanish Astronomical Centre, Calar Alto, (Spain) operated by the 
          Max-Planck-Institut f\"{u}r Astronomie and on observations obtained at the 
          Gemini Observatory, which is operated by the Association of Universities for 
          Research in Astronomy, Inc., under a cooperative agreement with the NSF on 
          behalf of the Gemini partnership: the NSF (United States), the PPARC 
          (United  Kingdom), the NRC (Canada), CONICYT (Chile), the ARC (Australia), 
          CNPq (Brazil) and CONICET (Argentina).}}

   \author{Tigran~A. Movsessian
          \inst{1}
          \and
          Tigran~Khanzadyan
          \inst{2,3}
          \and
          Colin~Aspin 
          \inst{4}
          \and
          Tigran~Yu. Magakian
          \inst{1}
          \and
          Tracy~Beck 
          \inst{4}
          \and 
          Alexei~Moiseev
          \inst{5}
          \and
          Michael~D. Smith
          \inst{6,7}
          \and
          Elena~H. Nikogossian
          \inst{1}
          }


   \institute{Byurakan Astrophysical Observatory, 378433 Aragatsotn reg., 
              Armenia\\
                \email{tigmov@web.am, tigmag@sci.am, elena@bao.sci.am}
          \and
              Centro de Astrof\'\i sica da Universidade do Porto, 
              Rua das Estrelas, 4150 -- 762 Porto, Portugal\\
               \email{khtig@astro.up.pt}
          \and
              Max-Planck Institut f\"{u}r Astronomie, K\"{o}nigstuhl 17,
              D-69117 Heidelberg, Germany
          \and
              Gemini Observatory, 670 N. Aohoku Place, Hilo, HI 96720 \\
              \email{caa@gemini.edu,tbeck@gemini.edu}
          \and
              Special Astrophysical Observatory,
              N.Arkhyz, Karachaevo-Cherkesia, 369167 Russia\\
             \email{moisav@sao.ru}
          \and
             Armagh Observatory, College Hill, Armagh BT61 9DG, 
             Northern Ireland, United Kingdom\\
                \email{mds@arm.ac.uk}     
          \and
             Centre for Astrophysics and Planetary Science,  
             University of Kent, CT2 7NR, United Kingdom
             }

   \date{Received ...; accepted 18/05/2006}

   \abstract
    {To follow the early evolution of stars we need to understand how young stars accrete 
     and eject mass. It is generally assumed that the FU~Orionis phenomenon is related to 
     the variations in the disk accretion, but many questions remain still open, in 
     particular because of the rarity of FU~Ori type stars.} 
    {We explore here the characteristics of the outburst and of the environment of one 
     new object, discovered recently in the in the active star formation region 
     containing RNO~127, within the Cygnus OB7 dark cloud complex.}
    {We present an extensive optical and near-infrared study of a new candidate of 
     FU~Orionis object, including its direct imaging, spectroscopy and scanning 
     Fabry-P\'erot interferometry.} 
    {The source, associated with the variable reflection nebula, underwent prodigious 
     outburst. The "Braid" nebula, which appeared in 2000, as is indicated by its name, 
     consists of two intertwined features, illuminated by the outburst. Subsequent NIR 
     observations revealed the bright source, which was not visible on 2MASS images, and
     its estimated brightening was more than 4 magnitudes. Optical and infrared spectral
     data show features, which are necessary for the system to be referred to as a 
     \textbf{FUor} object. The bipolar optical flow directed by the axis of nebula also 
     was found. Various estimates give the November/December 1999 as the most probable 
     date for the eruption.}
   {}

   \keywords{ISM -- Stars: formation --  ISM: jets and outflows -- 
  ISM: clouds}
   
 \titlerunning{An FUor Object in Cygnus OB7}
  \authorrunning{Movsessian et al.}   
   
   \maketitle

\section{Introduction}

\begin{table*}
 \caption{Log of the observations and used archive data.}
 \label{obs-log}
 \begin{tabular*}{\textwidth}{%
        @{}l@{\extracolsep{\fill}}%
        @{}c@{\extracolsep{\fill}}%
        @{}c@{\extracolsep{\fill}}%
        @{}c@{\extracolsep{\fill}}%
        @{}c@{\extracolsep{\fill}}%
        @{}c@{\extracolsep{\fill}}%
        @{}c@{\extracolsep{\fill}}%
        @{}r@{\extracolsep{\fill}}}
 \hline
 \hline
 \noalign{\smallskip}
   \sf{Date (UT)}
  &\sf{Band~Name}
  &\sf{Obs.~Mode}
  &\sf{Origin}
  &\sf{$\lambda_{c}$}
  &\sf{Pixel~scale($\arcsec$)}
  &\sf{Seeing($\arcsec$)}
  &\sf{Exposure(sec)}\\
\noalign{\smallskip}
\hline
\noalign{\smallskip}
24.Aug.1990&R        &Imaging    &POSS-II     &5400\AA\      &1.00&$\sim$2.0&4200           \\
21.Jun.1999&Ks       &Imaging    &2MASS       &2.17$\mu$m    &1.00&$\sim$3.5&   7.8         \\
22.Aug.2000&I        &Imaging    &2.6m        &8200\AA\      &0.65&$\sim$2.0& 600           \\
09.Dec.2000&Ks       &Imaging    &3.5m        &2.196$\mu$m   &0.39&$\sim$1.7& 710           \\
21.Apr.2001&I        &Imaging    &2.6m        &8200\AA\      &0.42&$\sim$2.0& 600           \\
07.Nov.2001&V        &Imaging    &2.6m        &5300\AA\      &0.42&$\sim$2.0& 600           \\
07.Nov.2001&R        &Imaging    &2.6m        &6400\AA\      &0.42&$\sim$2.0& 600           \\
07.Nov.2001&I        &Imaging    &2.6m        &8200\AA\      &0.42&$\sim$2.0& 600           \\
27.Nov.2002&H$\alpha$&Fabry-P\'erot&6.0m      &6560\AA\      &0.56&$\sim$1.0&  36$\times$180\\
22.Aug.2003&K        &Imaging    &Gemini North&2.2$\mu$m     &0.12&$\sim$0.6&  30           \\
26.Sep.2003&-        &Spectr.    &2.6m        &3900-7200\AA\ &0.42&$\sim$1.5&4800           \\
08.Sep.2004&-        &Spectr.    &6.0m        &6100-7200\AA\ &0.18&$\sim$3.0&4200           \\
06.Sep.2005&-        &Spectr.    &IRTF        &1.3-4.1$\mu$m &0.15&$\sim$0.8&1200           \\
\noalign{\smallskip}
\hline
\end{tabular*}
\end{table*}

There is now strong evidence that the FU~Orionis (FUor) outburst phenomenon is 
closely related to the earliest stages of stellar evolution 
\citep{2003ApJ...595..384H}. However, the mechanism by which these eruptions 
occur is still poorly understood although an integral link to accretion and the 
circumstellar accretion disk is established 
\citep{2004ApJ...606L.119R,2004ApJ...609..906H}. It is also clear that the 
creation of collimated outflows is intimately linked to the presence of 
accretion disks even though it is not yet understood whether FUor events trigger 
directed outflows. This link is difficult to establish given the rarity 
of known `classical' FUors  i.e. those that have been observed to display large 
brightness increases on short timescales coupled with significant changes in
spectral characteristics \citep{1977ApJ...217..693H}. FUors are
characterized by sudden brightness increases of 5--6 magnitudes on a
timescale of as little as 6 months to as long as 15 years.  After the
eruption, they start to fade on timescales of many decades to even
centuries.  Their optical spectra after outburst are composite in that
they display A-type stellar features in UV and blue, and F to G-type
stellar in red.  A very low surface gravity is implied by the optical
features present similar to those found in super-giant atmospheres.
They typically have a complex H$\alpha$ profiles with a P~Cygni shape
indicating fast outflowing neutral gas with velocities up to 1000
km~s$^{-1}$. In the near-IR their spectrum is dominated by late
M-type stellar features, in particular, deep CO band-heads.  They are
found to be associated with dark cloud complexes and exhibit prominent
reflection nebulae.

Several FUor-like objects have recently been discovered e.g. PP13\,S 
\citep{1998A&A...333.1016S}, and NGC~2264~AR~6ab \citep{2003AJ....126.2936A}. 
These are termed FUor-like since the original outburst and large brightness 
increase were not documented. At least some of these FUor-like objects are 
probably also FUors. Since it is unclear whether all FUor events possess
similar characteristics such as, for example, rise and decay timescales, and 
changes in photometric and spectroscopic signatures, more searches for new 
FUors are urgently required. In total, 15 FUor-like objects are currently known 
\citep{2005MmSAI..76..320V}. Only by quantifying these and other parameters 
over a statistical sample of objects will enable us to understand the 
nature of the eruptions and the resultant effects.

We present here an optical and near-infrared (NIR) investigation of a new
FUor-like outburst recently discovered in an active star formation
region surrounding RNO~127, located in the Cygnus OB7 dark cloud
complex \citep{2003A&A...412..147M} (henceforth MKMSN03). In this
region, several new cometary nebulae, Herbig-Haro objects and
outflows/jets were found (MKMSN03). The detection of a new
NIR reflection nebula in this region immediately
brought our attention to the possibility of its creation by a FUor
outburst. This nebula, designated `IR-Neb' in MKMSN03, has already
been described in detail in that paper. Here, we bring together all
available evidence on the nature of the optical and NIR nebulosity and
the underlying young star with the aim of determining whether the
associated young star has undergone a FUor outburst.  Before we
can proceed with the description of this object, we have decided to
name it based on its physical appearance - the ``Braid'' nebula. This
is to avoid confusion since the nebula was originally detected in the
optical and has many similarities with other known nebulae
(e.g. Chameleon IR Neb) also called ``IR-Neb''.

\section{Observations and analysis}

Table\,\ref{obs-log} provides the log of observed and archive data. The first column
lists the UT date of the observations, the second column gives the names of 
the band where applicable, and the third column indicates the observing mode 
(imaging or spectroscopy). The fourth column lists the origin of the data, the 
fifth column gives the central wavelength of the band or the spectral range in 
case of spectroscopy, the sixth column lists the pixel scale of the detector 
used while the seventh and eighth columns give the seeing and exposure values,
respectively.

\subsection{Optical investigations}
\subsubsection{Imaging}
\label{opt-img}

Direct optical images of the region containing RNO~127 (RA=21~00~31.8, 
Decl.=+52~29~17, J2000) were obtained with the 2.6m telescope of the
Byurakan Observatory, Armenia, during several epochs starting from August
2000 with the ByuFOSC-2 \citep{2000BaltA...9..652M} camera at prime focus. 
A Thomson 1060$\times$1028 CCD detector was used working in `half obscured 
mode' with a resultant image size of 1060$\times$514 pixel 
\citep[see][for mode explanation]{2000BaltA...9..652M}. The field of view was 
11.5\arcmin$\times$5.5\arcmin\ at a pixel size of 0.65\arcsec. From November 
2001 all observations were performed using 
the multi-mode SCORPIO{\footnote{It's brief description can be found 
via the link {\tt http://www.sao.ru/$\backsim$moisav/scorpio/scorpio.html}}} 
camera with a Lick 2063$\times$2058 pixel CCD detector. This camera provides 
a field of view of 14\arcmin$\times$14\arcmin\ with 0.42\arcsec\ pixel-defined
resolution. 

In addition to our direct imaging at the Byurakan Observatory, we have used 
the POSS-II{\footnote{The Second Palomar Observatory Sky Survey 
(POSS-II) was made by the California Institute of Technology with funds from 
the National Science Foundation, the National Geographic Society, the Sloan 
Foundation, the Samuel Oschin Foundation, and the Eastman Kodak Corporation.}}
r-band image of the same region for a comparative study. Details are summarized 
in Table\,\ref{obs-log}. Data reduction was described in our previous paper
(MKMSN03).

\begin{figure*}[]
 \centering
\mbox{
  \subfigure[DSS2 R]{\includegraphics[width=8.5cm]{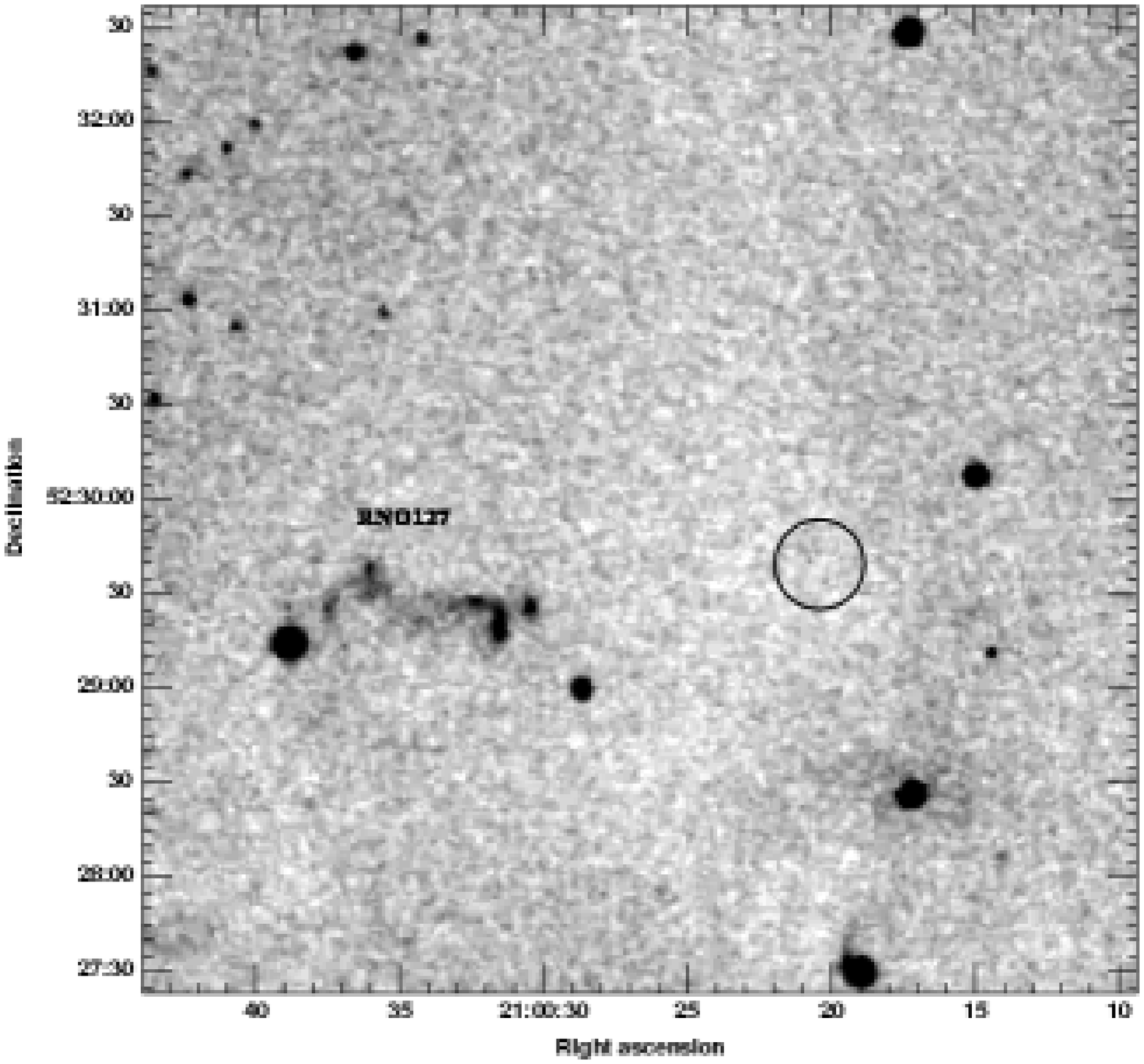}}\quad  
  \subfigure[Our R]{\includegraphics[width=8.5cm]{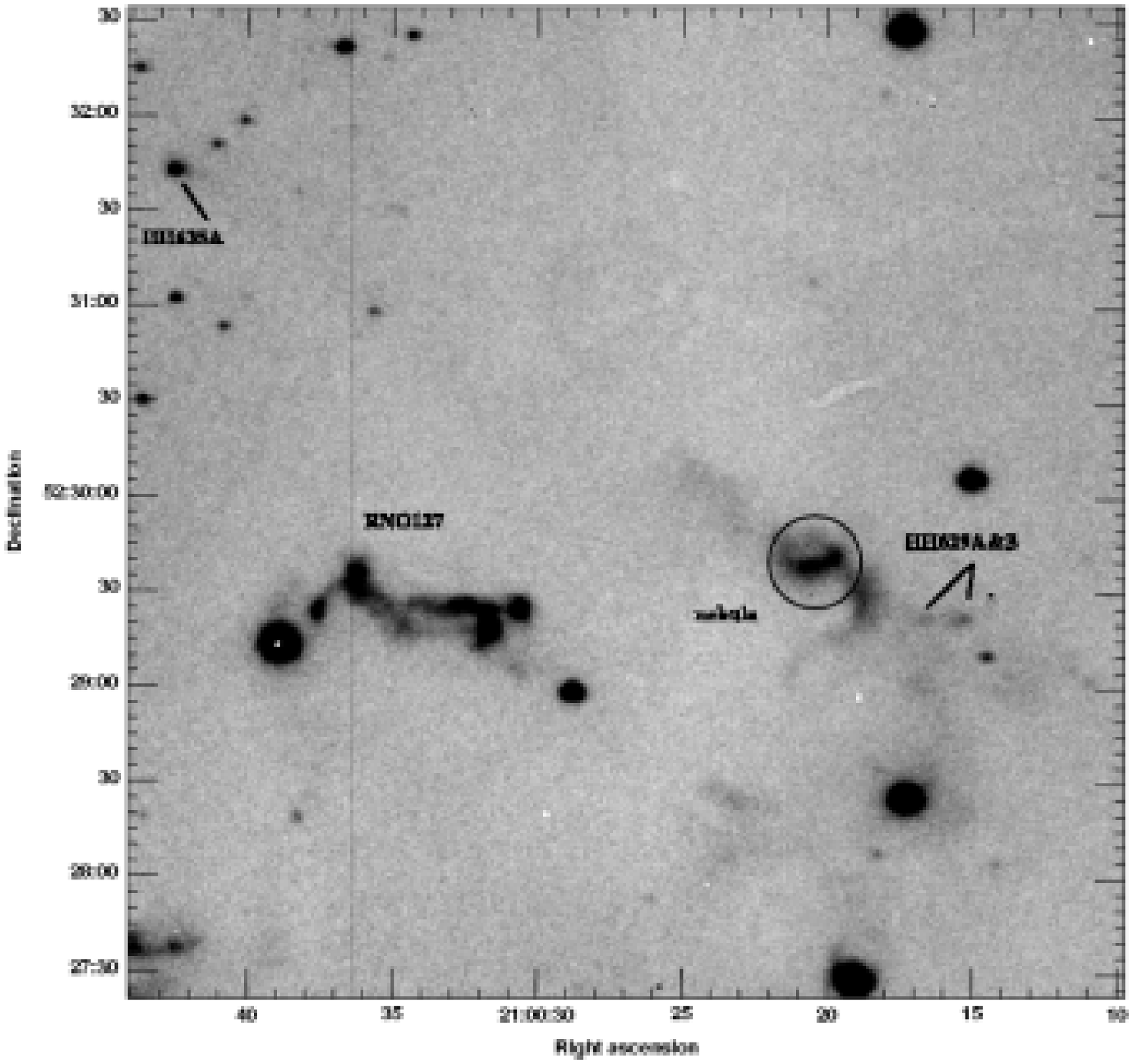}}\quad
}
 \caption{The region of Cygnus OB7 containing RNO127 and Braid nebula. 
          (a) This is an optical POSS-II R-band image taken 4th of August 1990 
          which does not show the optical nebulosity now associated with the 
          IR Nebula. The black circle indicates the place where the optical 
          nebula is now located. 
          (b) This is our CCD optical R-band image taken 19th of November 2001. 
          The brightness of RNO~127 in (a) and (b) suggests that the optical 
          nebulosity associated with the Braid Nebula should have been 
           detected in the POSS-II images if it were then visible.}
  \label{region}
\end{figure*}

\subsubsection{Spectroscopy}

Long-slit optical spectroscopy of the brightest section of the Braid nebula was
performed with the SCORPIO camera, equipped with the same CCD detector
as described \S\ref{opt-img}, on the 2.6m Byurakan Observatory telescope.
The disperser employed was a 600 lines/mm grism providing 10\AA\ resolution. The
long-slit was 2\arcsec\ wide and 6\arcmin\ long oriented EW.

In addition, a spectrum with higher spectral resolution was obtained on the
6m telescope of the Special Astrophysical Observatory (Russia) using a similar
SCORPIO device \citep{2005AstL...31..194A} equipped with a 2048$\times$2048
pixel EEV CCD detector. A dispersing holographic grism with 1800 lines/mm was
used, which gave a resolution of about 2.5\AA\ in the red spectral range. The
same brightest region in the Braid nebula was again observed.

Again dates, spectral range and observing conditions are listed in
Table\,\ref{obs-log}. Data reduction was performed in a standard manner using
routines written under IDL environment developed by V. Afanasiev 
\citep[see short description in][]{2005AstL...31..194A}. The standard star 
HD217086 was used to calibrate the spectra.

\subsubsection{Scanning Fabry-P\'erot interferometry}

Observations were carried out at the prime focus of the 6m SAO telescope (see
Table\,\ref{obs-log}). We used a scanning Fabry-P\'erot interferometer (IFP)
placed in the parallel telescope beam of the SCORPIO focal reducer. The
SCORPIO capabilities in IFP observations are described by
\citet{2002BSAO...54...74M}. The detector was a Tektronix 1024$\times$1024
pixel CCD array. The observations were performed with 2$\times$2 pixel binning
to reduce the readout time, and so 522 $\times$ 522 pixel images were obtained
in each spectral channel. A field-of-view of 4.8\arcmin\ square was observed
with a scale of 0.56\arcsec\ per pixel. An interference filter with FWHM=15\AA\
centered on the H$\alpha$ line was used for pre-monochromatization.

We used a Queensgate ET-50 interferometer operating in the 501st order at the
wavelength of H$\alpha$ which provided a spectral resolution of
FWHM$\approx$0.8\AA\ (equivalent to $\sim$40 kms$^{-1}$) for a range of
$\Delta\lambda$=13\AA\ ($\sim$ 590 km s$^{-1}$) free from order overlapping.
The number of spectral channel images obtained was 36 with the size of a
single channel being $\Delta\lambda\approx$0.36\AA\ ($\sim$ 16 km s$^{-1}$).

We reduced our interferometric observations by using the IDL-based software
developed at the SAO \citep{2002BSAO...54...74M}. After the primary data
reduction, the subtraction of night-sky lines, and wavelength calibration, the
observations were represented by a ``data cube'' in which each point in the
522$\times$522 pixel field contains a 36 channel spectrum. We performed
optimal data filtering e.g. Gaussian smoothing over the spectral coordinate
with a FWHM=1.5 channels and spatial smoothing by a two-dimensional Gaussian
with a FWHM=2--3 pixels. This was achieved using the ADHOC software 
package{\footnote{The ADHOC software package was developed by J. Boulestex
(Marseilles Observatory) and is publicly available from 
\tt{http://www-obs.cnrs-mrs.fr/adhoc/adhoc.html}.}}.

\begin{figure*}[]
 \centering
  \includegraphics[width=14cm]{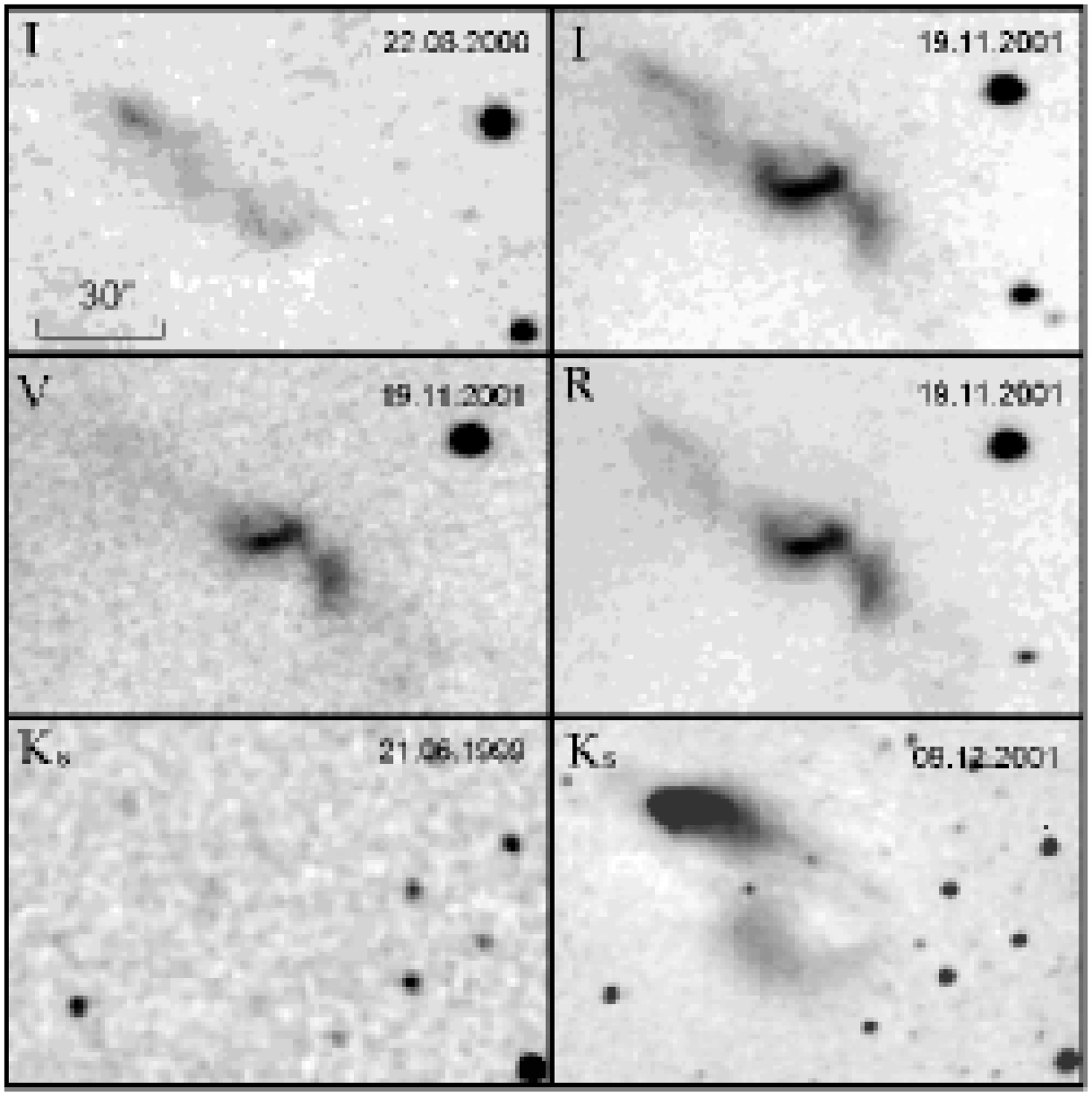}
  \caption{Images of the Braid nebula in the I band, obtained with the 2.6m telescope 
           at various epochs (upper row). The same nebula in V and R bands, 
           imaged with the 2.6m telescope (middle row). The appearance of the nebula 
           in the K$_{s}$ band from the 2MASS survey and  the 3.5m telescope image 
           (lower row). The dates of the observations are shown in the upper right 
           corners.}
  \label{all-figures}
\end{figure*} 

\subsection{Near-infrared exploration}

\subsubsection{Imaging}

For the NIR images we first used the Omega Prime \citep{1998SPIE.3354..825B} 
camera at the Calar Alto 3.5m telescope, Spain (see Table\,\ref{obs-log}). 
Observations were done using Full MPIA double correlated read mode 
\citep[see][for full description]{1998SPIE.3354..825B}, which has a 
minimum detector integration time (DIT) of 1.677\,s. By using this method 
each `individual frame' was constructed from 47 co-added DITs resulting in 
about 79\,s of integration. In this way, 9 frames were taken with a standard 5-point 
jittering. The data reduction and analysis are described in MKMSN03.

Additional NIR K band images of the Braid nebula were obtained at the 
``Fredrick C. Gillett" Gemini North Telescope on Mauna Kea, Hawaii 
(see Table\,\ref{obs-log}). The facility near-IR imager NIRI 
\citep{2003PASP..115.1388H} was used for the observations with a K-band filter. 
The total exposure time
for the images was 30 seconds. A small dither sequence was applied keeping
the IR object on the image in all spatial positions. These images
were taken to attempt to see high spatial resolution outflow structure
close to the bright IR object and compare the point-spread function of
the IR object to that of nearby stellar sources. Since the data were
not photometric (due to variable cloud cover), we reduced the image by
subtracting adjacent frames with the object located at different
positions on the detector. This yielded first-order flat-field
corrections and allowed us to study the morphology of the object in
detail.

Finally, we utilized the 2MASS Sky Atlas{\footnote{2MASS Sky Atlas 
Image obtained as part of the Two Micron All Sky Survey (2MASS), a joint 
project of the University of Massachusetts and the Infrared Processing 
and Analysis Center/California Institute of Technology, funded by the 
National Aeronautics and Space Administration and the National Science 
Foundation.}} to study the region at an earlier epoch (see Table\,\ref{obs-log}). 
The 2MASS images were used to calibrate the Calar Alto 
NIR images using several 2MASS Point Source Catalog objects in the field. 
Zero-points (ZP) were determined and magnitudes where assigned for the source 
using the aperture photometry tool in GAIA (see \S\ref{nirimphot} on used 
apertures).

\subsubsection{Spectroscopy}

The NIR spectroscopy of the stellar source associated with the nebula 
was obtained at the NASA Infrared Telescope Facility (IRTF) 3m Observatory 
on Mauna Kea, Hawaii (Table\,\ref{obs-log}). The facility infrared spectrograph 
``SpeX" \citep{2003PASP..115..362R} was used in the short and long wavelength 
cross-dispersed settings for R$\sim$1400 resolution spectra over the full 
1.3 - 4.1 $\mu$m region. The data were acquired using 7.5\arcsec\ beam-switched nods 
(ABBA) along the 15\arcsec\ slit. We used individual exposure times of 30s and 60s 
for the short and long wavelength modes, respectively. The nod pattern was 
repeated multiple times for total on-source integration times of 1200s in each 
setting. The facility data reduction package ``Spextool" \citep{2004PASP..116..362C} 
was used to combine, extract, telluric correct, merge the cross dispersed 
orders and clean the spectra for bad pixels. The Hipparchos A0 Star HIP 104127 
(K$_{s}$=7.51) was used for the telluric correction and flux calibration of 
the data; the photospheric hydrogen absorption features in this star were 
removed using the deconvolution technique incorporated into the Spextool 
package \citep{2003PASP..115..389V}.

\section{Results}

\subsection {Optical imaging and photometry}

In Fig.\ref{region}a we show the region of Cygnus OB7 containing 
RNO~127 and the Braid nebula. This image is from the Digital Sky Survey 
POSS-II plates and is in the R-band. The plate was taken on 
24 August 1990 and does not show any optical nebulosity associated 
with the Braid nebula. The location of the optical nebula seen in our more 
recent CCD images is marked in Fig.\ref{region}a by a black circle. 
Comparing the POSS-II plate to our R-band CCD image from MKMSN03 
(Fig.\ref{region}b) we see that the optical nebula is, at this time 
(19 November 2001), well detected. In fact, RNO~127 is visible in both 
the POSS-II plate and our CCD image (lower-left quadrant) and its detection 
in both suggests that if the optical nebulosity associated with the Braid 
nebula was present in 1990 at the same brightness as in 2001, it should 
have been detected.

Fig.\ref{all-figures} upper and middle panels show the extent of the 
optical nebulosity in the I, R and V broad bands. Our optical images, 
shown here and in MKMSN03, clearly show an approximately linear 
progression of nebulous knots extending from the location of the Braid nebula. Images in the same
passbands taken at different epochs after April 2001 do not show any
significant changes in the structure of the nebula nor in its
brightness. Upon closer inspection, the linear structure resembles a
series of curving or possibly helical filaments and, since it is seen
in all passbands, it should be mainly of a reflection origin.

We have attempted to estimate the optical colors of the nebula by measuring 
its surface brightness in the brightest region. Photometry in V, R and I bands 
was performed for the data obtained in 2002 and 2004, calibrated using 
the NGC7790 cluster (2002) and M92 cluster (2004). Very similar values were 
obtained in two epochs, i.e. V$-$R = 1.16; R$-$I = 1.02 in 2002 and 
V$-$R = 1.21; R$-$I = 1.06 in 2004. This is consistent with there being no 
major changes in nebula illumination/excitation after the initial brightening 
and between the two epochs. In principle, these colors could allow us to 
estimate the spectral type of the illuminating source but since we do not 
know {\bf i)} the amount of overlying dust extinction/reddening between the 
nebula and us (reddening the observed colors), {\bf ii)} the extinction/reddening 
between the source and the nebula (again, reddening the observed colors), 
and {\bf iii)} the real effect of scattering and emission in the nebula 
(probably making the observed colors more blue), any estimate is unreliable. 
We note, however, that the observed colors of the nebula would correspond 
to a K5--M1 dwarf star.

\begin{figure}
 \centering
  \includegraphics[width=8cm]{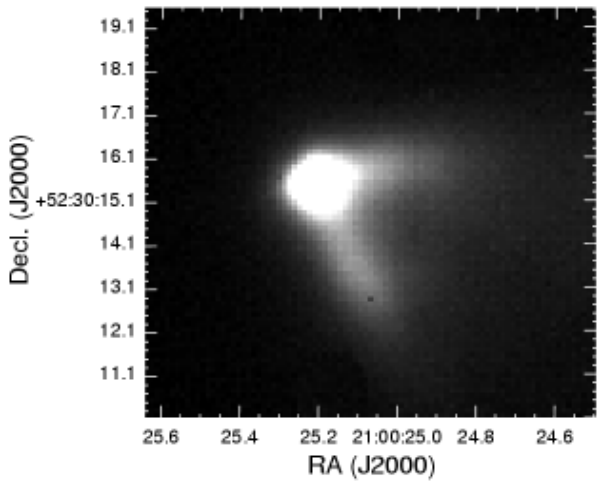}
  \caption{The region closest to the star-like source at the apex of the 
           Braid nebula. This is a K band image from Gemini using 
           NIRI. North is up, east is to the left. Note the extensions 
           suggesting we are detected the cavity walls of an outflow.}
  \label{irneb-hires}
\end{figure} 

\subsection {Optical spectroscopy}

\begin{figure}
 \centering
  \includegraphics[width=8.5cm]{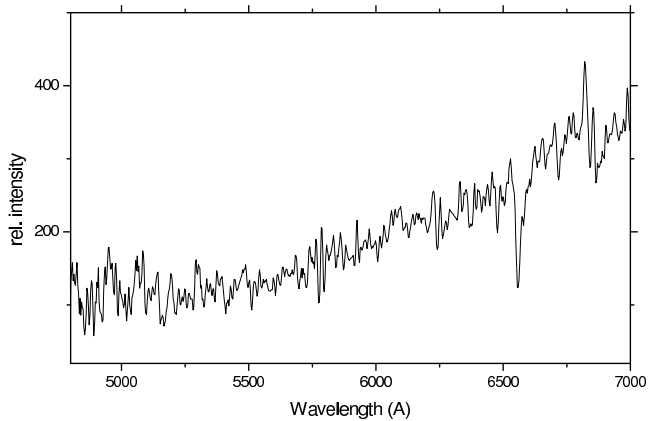}
  \caption{The low-resolution optical spectrum of the reflection nebula.}
  \label{lowres-spec}
\end{figure} 

\begin{figure}
 \centering
  \includegraphics[width=8.5cm]{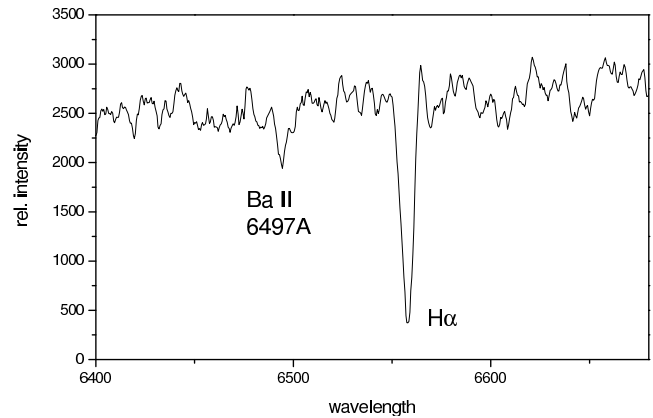}
  \caption{The H$\alpha$ line and the 6494\AA\ BaII blend in the optical spectrum of
           the reflection nebula.}
  \label{hires-spec}
\end{figure}

Our low spectral resolution optical spectrum of the brightest knot of
the optical nebula has a S/N$\sim$3 and only an H$\alpha$ absorption 
line can be identified (see Fig.\ref{lowres-spec}) in addition to the rising
red continuum. This makes the optical spectrum of Braid nebula similar to that
of FU~Orionis itself \citep{2006AJ..xx.xxxx} and another known FUor, BBW~76 
\citep{2002AJ....124.2194R}.

A higher spectral resolution spectrum (Fig.\ref{hires-spec}) confirms
the presence of these features in the spectrum of the optical
nebulosity. We detect strong and wide ($\sim$ 600 km~s$^{-1}$)
H$\alpha$ absorption with a small redshifted emission component. The
BaII absorption blend is also present.

One noteworthy point is the asymmetry in the H$\alpha$ absorption profile. 
It is clear that the blue wing of absorption is more extensive than the 
corresponding red wing. The radial velocity corresponding to the H$\alpha$ 
deepest absorption is $-220$ km~s$^{-1}$, and its blue wing can be traced 
out to $-540$ km~s$^{-1}$. Such a profile is usually considered as evidence 
of high mass loss in the form of a stellar wind and is a typical 
characteristic of FUors. The velocity of a probable emission component is low: 
$+20$~km~s$^{-1}$ and may correspond to ongoing and active accretion. On the whole, 
this profile is quite similar to the spectra of classic FUors 
\citep{2003ApJ...595..384H}.

\begin{figure}
 \centering
  \includegraphics[width=8cm]{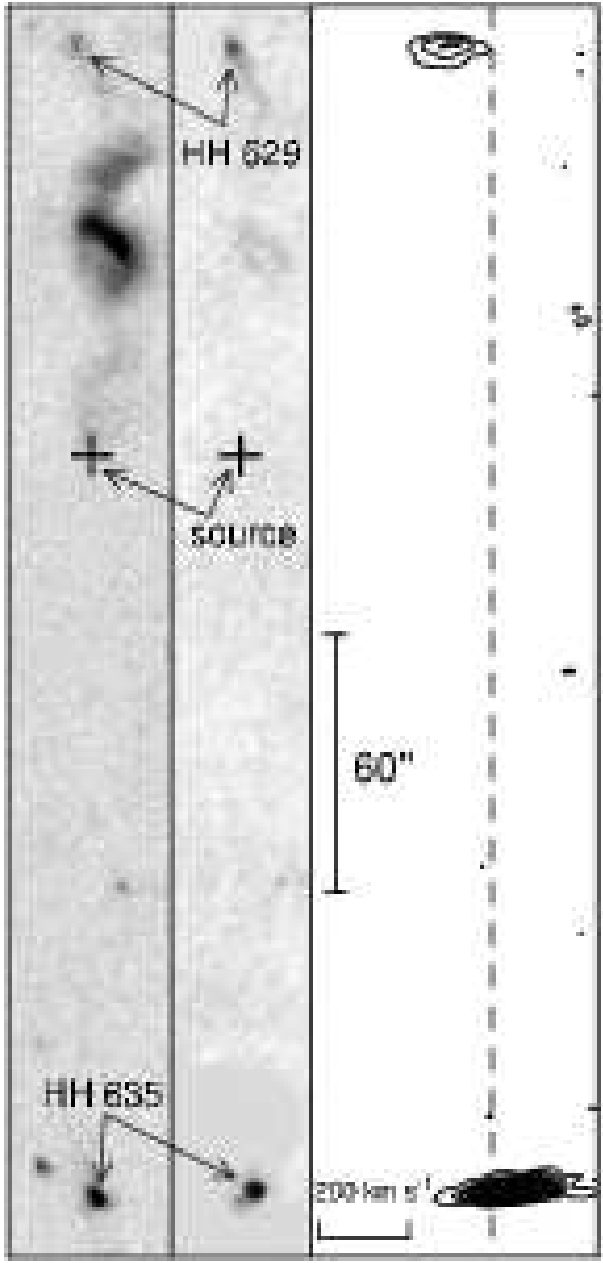}
  \caption{{\bf Left panel:} the [SII] image of the Braid nebula and HH-objects, oriented  
           along the axis of the nebula; position of the source is marked by the
           cross. {\bf Center panel:} the monochromatic image in H$\alpha$, obtained as
           the sum of all FP frames. {\bf Right panel:} PV diagram along the flow.}
  \label{pv}
\end{figure}


\subsection {Fabry-P\'erot interferometry} 


Two HH objects are aligned with the axis of the optical reflection
nebula: HH\,629 107\arcsec\ to the south-west
and HH\,635A 173\arcsec\ north-east of the Braid nebula (MKMSN03).  In [SII] emission,
HH\,629 is resolved into two close knots, A and B (also identified in Fig.\ref{region}) 
while in H$\alpha$ it appears more diffuse and extended.

MKMSN03 suggested that HH\,629 and HH\,635A, as well as several other more distant
faint HH objects, are associated with a bipolar outflow driven from
the protostar associated with the Braid nebula. This suggestion is
supported kinematically by our Fabry-P\'erot scanning interferometry of
this region. HH\,629 has a negative radial velocity of $-$92~km~s$^{-1}$ while 
HH\,635A has a positive velocity of +38~km~s$^{-1}$.
We show in Fig.\ref{pv} the monochromatic image of the region including 
the HH objects, which was extracted from the Fabry-P\'erot data. The 
position-velocity diagram along the HH flow, which clearly shows the 
bipolarity, is also presented in this figure. It should also be
noted that the redshifted HH\,635A has a significant velocity gradient: its 
leading edge has a radial velocity of only +16 km~s$^{-1}$ while its trailing 
edge reaches +73 km~s$^{-1}$. In contrast, HH\,629 has no such obvious velocity gradient.

The Fabry-P\'erot datacube also confirms that the optical nebula is
purely reflective in nature with no trace of any emission lines such
as H$\alpha$ emission along its extent.

\subsection{Near infrared imaging and photometry}
\label{nirimphot}

In the NIR Ks band image shown in Fig.\ref{all-figures}
(lower row, right panel) the source appears dramatically different
from the optical. At 2 microns we observe a cometary reflection
nebula with a very bright core. The nebula is $\sim$1\arcmin\ in total extent
and again exhibits a curving morphology.
 
Two years earlier, the same region was imaged as part of the 2MASS survey
and is displayed in the left panel of the lower row in Fig.\ref{all-figures}.
Clearly, dramatic changes have occurred in the object at NIR
wavelengths. We also note that it is possible to discern a faint
counter-fan structure, extending to the north-east from the brightest
point of the nebula (see lower row, right panel of the
Fig.\ref{all-figures}).

Photometry of the Braid nebula was also performed. In 1999, the object is 
barely visible and has an m$_k$=15.1 ($\pm$0.5). In 2001, $\sim$17 months 
later, the object has an m$_k$=10.45 ($\pm$0.5). Both values were measured 
in the same elliptical aperture of $\sim$40$\arcsec\times$20$\arcsec$. 
Hence, we have observed a near-IR brightening of at least 4 magnitudes.

If we look at the star-like object at higher spatial resolution from our 
Gemini NIRI image (Fig.\ref{irneb-hires}), we see that at
low contrast the star is resolved into a point-like source with linear
extensions protruding to the south-west and west. We interpret this as direct 
detection of inner regions of the outflow cavity from the young star. The 
full-width half-maximum (FWHM) of the stellar profile
is $\sim$0.8\arcsec\ in this image while other stellar sources in the region
have a FWHM of $\sim$0.55\arcsec.


\subsection{Near-infrared spectroscopy}

\begin{figure}
 \centering
  \includegraphics[angle=90,width=8.5cm]{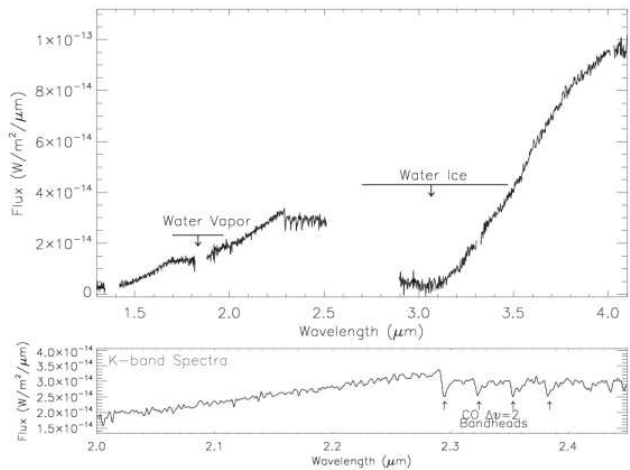}
  \caption{{\bf Upper panel:} the 1.3-4.1$\mu$m spectrum of the IR source with the water vapor
           and water-ice features identified. {\bf Lower panel:} the K-band 
           2.0\,--\,2.45\,$\mu$m spectrum showing the strong CO $\Delta${\it v}=2 
           bandhead absorption.}
  \label{irspec}
\end{figure}

The full 1.3\,--\,4.1\,$\mu$m spectrum of the NIR point source associated 
with the nebulosity is displayed in the upper panel of Fig.\,\ref{irspec},
and a zoomed view of the K-band spectrum (2.0\,--\,2.45\,$\mu$m) is shown in the 
lower panel. The spectrum shows steeply rising continuum emission 
in the 1\,--\,4\,$\mu$m region. No atomic or molecular absorption features 
characteristic of the photospheres of young stars are detected, but the 
shape of the NIR spectrum is affected by absorption features that probably 
arise from material in a disk. This supports our suggestion from \S\ref{nirimphot}
that we are not seeing the stellar photosphere even at 2\,$\mu$m. 
Weak water vapor absorption is detected in the 
1.8\,--\,2.0\,$\mu$m region, and strong CO $\Delta${\it v}=2 bandhead absorption 
is detected at 2.28\,--\,2.38\,$\mu$m. Although they are typically found in cool 
stars with low photospheric temperatures, water vapor features are also seen 
in the infrared spectra of many FUors \citep{1996AJ....112.2184G}. 
Additionally, such deep CO bandhead absorption arises from warm gas in the 
inner regions of circumstellar disks and is often a characteristic of variable 
young stars \citep{1997AJ....114.2700R}. We conclude, therefore, that the emission
from the inner regions of the circumstellar disk around this young star probably 
dominates the flux output in the NIR.

The continuum level of the IR source at wavelengths longward of $\sim$\,2.8\,$\mu$m 
is strongly affected by the broad water ice absorption centered at 
$\sim$\,3.05\,$\mu$m. This feature arises from the stretch-mode oscillation of O-H 
bonds in water molecules that are frozen onto dust grains in interstellar and 
circumstellar environments \citep{1975MNRAS.173..489C,1988MNRAS.233..321W}. 
The shape of the absorption profile is known to be temperature dependent, and 
a broad, rounded water feature as seen in our spectrum suggests that the ices 
are in a cold environment with T$<$80K. The water absorption feature seen toward 
the IR source could arise from: {\bf i)} - ices on grains in cold material in an 
extended circumstellar disk or envelope encircling the central star, {\bf ii)} - from 
the ambient cloud material, or {\bf iii)} - from a combination of both. Regardless of 
the origin of the ice absorption, the presence of the feature reveals that a 
significant amount of dust and ice obscures our line of sight to the star.

\section{Discussion}

We conclude that the variable optical
nebula is illuminated by the source associated with the bright
NIR object. NIR photometry has additionally discovered that the
source has undergone an increase in brightness of over 4 magnitudes
between pre- and post-outburst phases.  We did not directly observe the
process of the brightening of the object. Nevertheless, we can
obtain an estimate of the time of the outburst using a light
propagation method. The length of the optical nebula in April 2001 was
84\arcsec\ while at the earlier date of August 2000 it was only
61\arcsec. After April 2001 (i.e. 19th of November 2001 in Fig.\ref{all-figures}), there were
no further changes in the length of the nebula.  This implies that
radiation from the illuminating source had already traversed
the available physical dust structure.  This suggests that the outburst
radiation traversed 23\arcsec\ in less than 200 days. Extrapolating back
to the Braid nebula star-like object then indicates that the outburst
would have occurred a maximum of an additional $\sim$ 530 days
earlier, i.e. in or after January 1999. This implies that the outburst
could have occurred before the 2MASS non-detection but, more
plausibly, that the radiation traversed the scattering jet structure
in considerably less time. Unfortunately we do not have observations
between August 2000 and April 2001 to set a more precise time when the
propagation terminated.

Another constraint can be placed by assuming that the brightening of
the object took place after the 21st of June 1999 (the 2MASS observation date)
and that the light-wave travelled the angular distance, $\theta$, of
61\arcsec\ in under $t_o = 428$ days. Given a distance D to the Braid
nebula, then the inclination angle of the jet structure to the plane
of the sky is $\cos^{-1} \theta~D/(c t_o)$.  In MKMSN03 it is stated that
the distance of this star formation complex can be 800~pc although this
estimate is rather uncertain. If we assume this distance for the
nebula, then either its inclination angle is about 48\degr\ (which is
definitely too large value) or the outburst took place more
recently than the 21st of June 1999 (which is much more probable).

In addition, we can make estimations using the kinematical properties 
of the bipolar optical HH-flow associated with the Braid nebula.
If we assume that the real velocity of the HH outflow is
about 300~km~s$^{-1}$ (in accordance with the velocity of the
absorption component of the P~Cyg type profile of the H$\alpha$ line 
(Fig.\,\ref{hires-spec}) and the mean observed absolute value of radial 
velocity is 70~km~s$^{-1}$, then the angle to the plane of the sky will be only 
13\degr.

\begin{figure*}
 \centering
  \includegraphics[width=12cm]{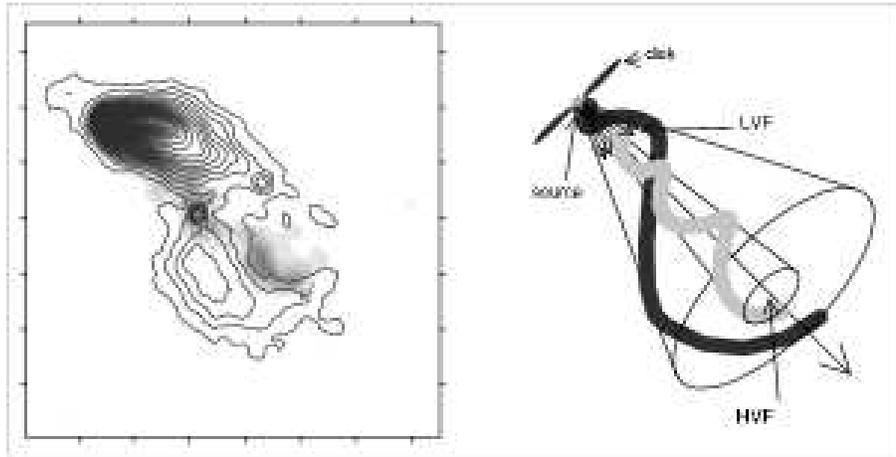}
  \caption{Image of the nebula in the K band (contours) and in the I band 
           (grayscale) (left panel), and the scheme of formation of double 
           helical structure on the surfaces of high and low velocity flows 
           (right panel).}
  \label{helix}
\end{figure*} 

Now we can assume that our estimates of the distance and inclination of 
the Braid nebula do not contradict the available data. In this case, 
assuming 800 pc and about 20\degr\ for these values, one can easily 
calculate that if in August 2000 the length of the visible nebula 
was about 61\arcsec, the light should have travelled this distance in 
approximately 290 days, i.e. the outburst most likely occurred in 
November/December 1999. 

An interesting feature of the reflection nebula, already mentioned in 
MKMSN03, is its quite different appearance in the optical and NIR ranges. 
Morphologically, both images have some indications of spiral structure but
with different opening angle and step (i.e. the physical separation between 
successive loops). To our best knowledge this is the first such example; 
however, there are other cases, when we see probable helical arms on the 
walls of the nebular cones. One of the best illustrations is the RNO~124 
nebula \citep{2004A&A...413..203M}, where two helical arms were detected.
Among other examples one can remember some evidences of the spiral structure 
in the famous NGC~2261 nebula \citep{1989MNRAS.239..665L, 1997ApJ...489..210C}.

It can be suggested that the wider arm of the Braid nebula, seen in NIR, 
represents the dust illuminated by outer parts of the circumstellar disk, 
while the narrow one reflects the direct stellar light. Their shape poses 
another problem. As was already mentioned in \S\ref{nirimphot}, it is very 
probable that there exists a cavity near the star (its opening angle is 
about 60\degr), created by the outflowing material along the axis of the 
nebula. If we indeed see two helices enclosing one another, then the pure 
schematic presentation of this structure can look like in Fig.\ref{helix}. 
The formation of helical ``density waves'' on the dust walls can be, perhaps, 
related with the instabilities on the interface surface between the flow 
and the surrounding medium. The formation of two helices can be understood 
if two components with significantly different velocities exist in the 
outflow; each component would be responsible for the formation of a distinct 
helix. Of course, these are only the preliminary speculations which need 
to be tested by modelling and observations.

Another significant point is that the Braid nebula source represents a still 
rare case of the FUor phenomenon with an optically detected outburst, which 
also is a source of a bipolar outflow. As in other similar cases: e.g., V346 Nor 
\citep{1985ApJ...289..331G} and probable FUors Z CMa
\citep{1989ApJ...338.1001H} and Re 50 \citep{1989A&A...220..249R}, the
kinematical age of the outflow (more than 2000 years for the distance
between HH\,635A and the source) is much greater than the time of the 
outburst. In fact, the kinematical age can be two times greater if more distant HH\,635B
and HH\,635C also belong to this outflow. 
This brings two possibilities: either each FUor outburst
adds new HH condensations into the flow (thus such outbursts must be
recurrent), or these two phenomena are not directly related.

To sum up, it may be said that the brightening of the object together with
its optical and NIR spectrum leads to the conclusion that the
illuminating star of the Braid nebula could well be another example
of the rare class of eruptive objects named FUors.

\begin{acknowledgements}

The authors wish to thank the anonymous referee for comments which substantially 
improved this manuscript. This work was mainly supported by INTAS grants 00-00287 
and 03-51-4838, and by grant of CRDF/NFSAT AS 062-02/CRDF 12009. T.Yu.M. and T.A.M. 
thank the administration of Gemini Observatory for the support of their visits in 
2003 and 2005. We also thank V.Afanasiev for obtaining the spectrum of Braid Nebula with 
6m telescope. We would like to acknowledge the data analysis facilities provided by the
Starlink Project which is run by CCLRC\,/\,Rutherford Appleton Laboratory on
behalf of PPARC. In addition, the following Starlink packages have been used:
CCDPACK, KAPPA and GAIA.

\end{acknowledgements}

\bibliographystyle{aa}
\bibliography{4609}

\begin{thebibliography}{29}
\expandafter\ifx\csname natexlab\endcsname\relax\def\natexlab#1{#1}\fi

\bibitem[{{Afanasiev} \& {Moiseev}(2005)}]{2005AstL...31..194A}
{Afanasiev}, V.~L. \& {Moiseev}, A.~V. 2005, Astronomy Letters, 31, 194

\bibitem[{{Aspin} \& {Reipurth}(2003)}]{2003AJ....126.2936A}
{Aspin}, C. \& {Reipurth}, B. 2003, \aj, 126, 2936

\bibitem[{{Bizenberger} {et~al.}(1998){Bizenberger}, {McCaughrean}, {Birk},
  {Thompson}, \& {Storz}}]{1998SPIE.3354..825B}
{Bizenberger}, P., {McCaughrean}, M.~J., {Birk}, C., {Thompson}, D., \&
  {Storz}, C. 1998, in Proc. SPIE Vol. 3354, p. 825-832, Infrared Astronomical
  Instrumentation, Albert M. Fowler; Ed., 825--832

\bibitem[{{Close} {et~al.}(1997){Close}, {Roddier}, {Hora}, {Graves},
  {Northcott}, {Roddier}, {Hoffmann}, {Dayal}, {Fazio}, \&
  {Deutsch}}]{1997ApJ...489..210C}
{Close}, L.~M., {Roddier}, F., {Hora}, J.~L., {et~al.} 1997, \apj, 489, 210

\bibitem[{{Cohen}(1975)}]{1975MNRAS.173..489C}
{Cohen}, M. 1975, \mnras, 173, 489

\bibitem[{{Cushing} {et~al.}(2004){Cushing}, {Vacca}, \&
  {Rayner}}]{2004PASP..116..362C}
{Cushing}, M.~C., {Vacca}, W.~D., \& {Rayner}, J.~T. 2004, \pasp, 116, 362

\bibitem[{{Graham} \& {Frogel}(1985)}]{1985ApJ...289..331G}
{Graham}, J.~A. \& {Frogel}, J.~A. 1985, \apj, 289, 331

\bibitem[{{Greene} \& {Lada}(1996)}]{1996AJ....112.2184G}
{Greene}, T.~P. \& {Lada}, C.~J. 1996, \aj, 112, 2184

\bibitem[{{Hartmann} {et~al.}(2004){Hartmann}, {Hinkle}, \&
  {Calvet}}]{2004ApJ...609..906H}
{Hartmann}, L., {Hinkle}, K., \& {Calvet}, N. 2004, \apj, 609, 906

\bibitem[{{Hartmann} {et~al.}(1989){Hartmann}, {Kenyon}, {Hewett}, {Edwards},
  {Strom}, {Strom}, \& {Stauffer}}]{1989ApJ...338.1001H}
{Hartmann}, L., {Kenyon}, S.~J., {Hewett}, R., {et~al.} 1989, \apj, 338, 1001

\bibitem[{{Herbig}(1977)}]{1977ApJ...217..693H}
{Herbig}, G.~H. 1977, \apj, 217, 693

\bibitem[{{Herbig} {et~al.}(2003){Herbig}, {Petrov}, \&
  {Duemmler}}]{2003ApJ...595..384H}
{Herbig}, G.~H., {Petrov}, P.~P., \& {Duemmler}, R. 2003, \apj, 595, 384

\bibitem[{{Hodapp} {et~al.}(2003){Hodapp}, {Jensen}, {Irwin}, {Yamada},
  {Chung}, {Fletcher}, {Robertson}, {Hora}, {Simons}, {Mays}, {Nolan}, {Bec},
  {Merrill}, \& {Fowler}}]{2003PASP..115.1388H}
{Hodapp}, K.~W., {Jensen}, J.~B., {Irwin}, E.~M., {et~al.} 2003, \pasp, 115,
  1388

\bibitem[{{Lightfoot}(1989)}]{1989MNRAS.239..665L}
{Lightfoot}, J.~F. 1989, \mnras, 239, 665

\bibitem[{{Lis} {et~al.}(1999){Lis}, {Menten}, \&
  {Zylka}}]{1999ApJ...527..856L}
{Lis}, D.~C., {Menten}, K.~M., \& {Zylka}, R. 1999, \apj, 527, 856

\bibitem[{{Moiseev}(2002)}]{2002BSAO...54...74M}
{Moiseev}, A.~V. 2002, Bull.~Special Astrophys.~Obs., 54, 74

\bibitem[{{Movsessian} {et~al.}(2000){Movsessian}, {Boulesteix}, {Gach}, \&
  {Zaratsian}}]{2000BaltA...9..652M}
{Movsessian}, T., {Boulesteix}, J., {Gach}, J.-L., \& {Zaratsian}, S. 2000,
  Baltic Astronomy, 9, 652

\bibitem[{{Movsessian} {et~al.}(2003){Movsessian}, {Khanzadyan}, {Magakian},
  {Smith}, \& {Nikogosian}}]{2003A&A...412..147M}
{Movsessian}, T., {Khanzadyan}, T., {Magakian}, T., {Smith}, M.~D., \&
  {Nikogosian}, E. 2003, \aap, 412, 147

\bibitem[{{Movsessian} {et~al.}(2004){Movsessian}, {Magakian}, {Boulesteix}, \&
  {Amram}}]{2004A&A...413..203M}
{Movsessian}, T.~A., {Magakian}, T.~Y., {Boulesteix}, J., \& {Amram}, P. 2004,
  \aap, 413, 203

\bibitem[{{Rayner} {et~al.}(2003){Rayner}, {Toomey}, {Onaka}, {Denault},
  {Stahlberger}, {Vacca}, {Cushing}, \& {Wang}}]{2003PASP..115..362R}
{Rayner}, J.~T., {Toomey}, D.~W., {Onaka}, P.~M., {et~al.} 2003, \pasp, 115,
  362

\bibitem[{{Reipurth}(1989)}]{1989A&A...220..249R}
{Reipurth}, B. 1989, \aap, 220, 249

\bibitem[{{Reipurth} \& {Aspin}(1997)}]{1997AJ....114.2700R}
{Reipurth}, B. \& {Aspin}, C. 1997, \aj, 114, 2700

\bibitem[{{Reipurth} \& {Aspin}(2004)}]{2004ApJ...606L.119R}
{Reipurth}, B. \& {Aspin}, C. 2004, \apjl, 606, L119

\bibitem[{{Reipurth} {et~al.}(2006){Reipurth}, {Aspin}, {Herbig}, {Beck},
  {Brogan}, \& {Connelley}}]{2006AJ..xx.xxxx}
{Reipurth}, B., {Aspin}, C., {Herbig}, G., {et~al.} 2006, in prep.

\bibitem[{{Reipurth} {et~al.}(2002){Reipurth}, {Hartmann}, {Kenyon}, {Smette},
  \& {Bouchet}}]{2002AJ....124.2194R}
{Reipurth}, B., {Hartmann}, L., {Kenyon}, S.~J., {Smette}, A., \& {Bouchet}, P.
  2002, \aj, 124, 2194

\bibitem[{{Sandell} \& {Aspin}(1998)}]{1998A&A...333.1016S}
{Sandell}, G. \& {Aspin}, C. 1998, \aap, 333, 1016

\bibitem[{{Vacca} {et~al.}(2003){Vacca}, {Cushing}, \&
  {Rayner}}]{2003PASP..115..389V}
{Vacca}, W.~D., {Cushing}, M.~C., \& {Rayner}, J.~T. 2003, \pasp, 115, 389

\bibitem[{{Vittone} \& {Errico}(2005)}]{2005MmSAI..76..320V}
{Vittone}, A.~A. \& {Errico}, L. 2005, Memorie della Societa Astronomica
  Italiana, 76, 320

\bibitem[{{Whittet} {et~al.}(1988){Whittet}, {Bode}, {Longmore}, {Admason},
  {McFadzean}, {Aitken}, \& {Roche}}]{1988MNRAS.233..321W}
{Whittet}, D.~C.~B., {Bode}, M.~F., {Longmore}, A.~J., {et~al.} 1988, \mnras,
  233, 321

\end{thebibliography}

\end{document}